\documentclass[graybox]{svmult}

\usepackage{graphicx}        % standard LaTeX graphics tool
\usepackage{amssymb}
\usepackage{amsmath}
\usepackage{amsfonts}
\usepackage{multicol}

\usepackage{makeidx}

\newcommand{\RR}{\mathbb{R}}

\newcommand{\CC}{\mathbb{C}}

\newcommand{\ZZ}{\mathbb{Z}}

\usepackage{stmaryrd}

\newtheorem{assumption}[theorem]{Assumption}

\newcommand{\UU}{\mathbf{U}}
\newcommand{\VV}{\mathbf{V}}

\def\ee{{(11)}}
\def\et{{(12)}}
\def\te{{(21)}}
\def\tt{{(22)}}

\begin{document}

\mainmatter

\title*{Eigenvalue based algorithms and software for the design of fixed-order stabilizing controllers for interconnected systems with time-delays}
 \titlerunning{Design of fixed-order stabilizing controllers}

\author{Wim Michiels and Suat Gumussoy}
 \institute{Wim Michiels \at Department of Computer Science, KU Leuven, Celestijnenlaan 200A, 3001 Heverlee, Belgium, \email{Wim.Michiels@cs.kuleuven.be}
 \and
 Suat Gumussoy \at MathWorks,  3 Apple Hill Drive, Natick, MA, 01760, USA,\ email{Suat.Gumussoy@mathworks.com
 \and
 This work is based on and the extension of the Authors' conference paper \cite{MichielsIFACTDS2012}.
 }
}
\authorrunning{W. Michiels and S. Gumussoy}

\maketitle

\abstract{
An eigenvalue based framework is developed for the stability analysis and stabilization of coupled systems with time-delays, which are naturally described by  delay differential algebraic equations.  The spectral properties of these equations are analyzed and their stability properties are studied, taking into account the effect of small delay perturbations.  Subsequently, numerical methods for stability assessment and for designing stabilizing controllers with a prescribed structure or order, based on a direct optimization approach, are briefly addressed. The effectiveness of the approach is illustrated with a software demo. The paper concludes by pointing out the similarities with the computation and optimization of $\mathcal{H}_{\infty}$ norms.
}

\section{Introduction}

We consider the stability analysis and stabilization of systems described by delay differential algebraic equations (DDAEs), also called descriptor systems~\cite{fridman}, of the form
\begin{equation}\label{wm:system}
%\left\{\begin{array}{l}
E \dot x(t)= A_0 x(t)+\sum_{i=1}^m A_i x(t-\tau_i),\ \  x(t)\in\RR^n,
%\end{array}\right.
\end{equation}
where $E$ is allowed to be singular.
The time-delays $\tau_i, i=1,\ldots,m$, satisfy
\[
0<\tau_1<\tau_2<\ldots<\tau_m
\]
and the capital letters are real-valued matrices of appropriate dimensions.

The motivation for the system description (\ref{wm:system}) in the context of designing controllers lies in its generality in modeling interconnected systems. For instance, the feedback interconnection of the system
\begin{equation}\label{wm:motex1}
\left\{\begin{array}{l}
\dot z(t) =\sum F_i z(t-r_i) + \sum G_i u(t-r_i)\\
y(t)=\sum H_i x(t-r_i) +\sum L_i u(t-r_i)
\end{array}\right.
\end{equation}
and the controller
\begin{equation}\label{wm:motex2}
\left\{\begin{array}{l}
\dot z_c(t) =\sum \hat F_i z_c(t-s_i) + \sum \hat G_i y(t-s_i) \\
u(t)=\sum \hat H_i z_c(t-s_i) +\sum \hat L_i y(t-s_i)
\end{array}\right.
\end{equation}
can be directly brought in the form (\ref{wm:system}), where
\[
x=[z^T\ z_c^T\ u^T\ y^T],\ \ \{\tau_1,\ldots,\tau_m\}=\{r_i\}\cup\{s_i\}.
 \]
 In this way no elimination of inputs and outputs is required, which may even not be possible in the presence of delays~\cite{reporthinf}.
 Another favorable property is the linear dependence of the matrices of the closed-loop system on the elements of the matrices of the controller.
 The increase in the number of equations, on the contrary, is a minor problem in most applications because the delay difference equations or algebraic constraints are related to inputs and outputs, as illustrated above, and
the number of inputs and outputs is usually much smaller than the number of state variables.
Finally, we note that also neutral systems can be dealt with in this framework, by introducing slack variables.   The neutral equation
\begin{equation}\label{wm:neutral1}
\frac{d}{dt} \left(z(t)+\sum_{i=1}^m G_i z(t-\tau_i)\right)=\sum_{i=0}^m H_i z(t-\tau_i)
\end{equation}
can namely be rewritten as
\begin{equation} \label{wm:neutral2}
\left\{\begin{array}{lll}
\dot{v}(t)&=&\sum_{i=0}^m H_i z(t-\tau_i) \\
0&=&-v(t)+ z(t)+\sum_{i=1}^m G_i z(t-\tau_i)
\end{array}\right.,
%
%\left[\begin{array}{cc}
%I & 0\\
%0&0
%\end{array}\right]
%\left[\begin{array}{c}
%\dot v(t)\\
%\dot z(t)
%\end{array}\right]=
%%
%\left[\begin{array}{cc}
%0 & H\\
%-I&I
%\end{array}\right]
%\left[\begin{array}{c}
% v(t)\\
% z(t)
%\end{array}\right]+
%\left[\begin{array}{cc}
%0 & 0\\
%0&G
%\end{array}\right]
%\left[\begin{array}{c}
% v(t-\tau)\\
% z(t-\tau)
%\end{array}\right],
\end{equation}
where $v$ is the slack variable. Clearly (\ref{wm:neutral2}) is of the form (\ref{wm:system}), if we set $x(t)=[v(t)^T\ z(t)^T]^T$.

The stability analysis of the null solution of (\ref{wm:system}) in this work is based on a spectrum determined growth property of the solutions, which allows us to infer stability information from the location of the characteristic roots. For instance,~exponential stability will be related to a strictly negative spectral abscissa (the supremum of the real parts of the characteristic roots). As we shall see, the spectral abscissa of (\ref{wm:system}) may not be a continuous function of the delays. Moreover, this may lead to a situation where infinitesimal delay perturbations destabilize an exponentially stable system. These properties are similar to the spectral properties of neutral equations. Since in a practical control design the robustness of stability against infinitesimal changes of parameters is a prerequisite, we will define the concept of strong stability, inspired by the common terminology  for neutral equations~\cite{have:02}, and we will introduce the notion of the robust spectral abscissa, which explicitly takes   small parametric perturbations into account. We will also provide explicit conditions and expressions that eventually lead to numerical algorithms.

%In order to compute the rightmost characteristic roots of (\ref{wm:system}), we combine a spectral discretization  of an infinite-dimensional linear system equivalent to (\ref{wm:system}), inspired by~\cite{breda:nonlocal,breda}, with local corrections of characteristic roots by Newton's method.

%Such a two-step approach, inferred from the dual interpretation of the eigenvalue problem associated  with a linear time-delay system as either infinite-dimensional and linear or finite-dimensional and nonlinear, also lies at the basis of the stability routine for steady state solutions of the package DDE-BIFTOOL~\cite{DDE-biftool} and of the method for computing $\mathcal{H}_{\infty}$ norms presented in~\cite{wimsimax}.

 Numerical algorithms for the computation of characteristic roots and the robust spectral abscissa are outlined, and  subsequently applied to the design of stabilizing controllers. Similarly to~\cite{joris}, a direct optimization approach towards stabilization is taken, based on minimizing the (robust) spectral abscissa as a function of the parameters of the controller. In the example (\ref{wm:motex1})-(\ref{wm:motex2}) these parameters may correspond to elements of the controller matrices. In this way stabilization is achieved on the moment that the objective function becomes strictly negative. This approach allows us to design stabilizing controllers with a prescribed structure or order (dimension). It is also possible to fix elements of the controller matrices, allowing to impose additional structure, e.g.,~ a proportional-integral-derivative (PID)-like structure, or sparsity.
%

%In the context of stability optimization of linear time-invariant (LTI) systems it is well known that the spectral abscissa is in general a nonconvex  function of the elements of the system matrices. In addition, it is typically not everywhere differentiable, even not everywhere Lipschitz continuous,  although it is differentiable almost everywhere~\cite{buleov:02,suatHIFOO}. These properties carry over to the case of the robust spectral abscissa of DDAEs under consideration. Therefore, special optimization methods for nonsmooth problems will be used, more precisely, the same methods underlying the package HIFOO for fixed-order $\mathcal{H}_{\infty}$ control design for LTI systems \cite{suatHIFOO}. The main differences with the stabilization routine in HIFOO are i) the \emph{infinite}-dimensional plant is controlled by a finite-dimensional controller, and ii) the potential sensitivity of the spectral abscissa w.r.t. infinitesimal perturbations of system parameters (delays), which must be taken into account.

  After a software demo of the stabilization algorithms we point out how the computational and optimization of $\mathcal{H}_{\infty}$ norm leads to similar problems as well as similar solutions and algorithms.

\vspace{-0.5cm}
\section{Preliminaries and assumptions}\label{wm:sectionprelim}
Let matrix $E$ in (\ref{wm:system}) satisfy
\[
\mathrm{rank}(E)=n-\nu,
\]
with $1\leq \nu<n$, and let the columns
of matrix $U\in\RR^{n\times \nu}$, respectively $V\in\RR^{n\times \nu}$, be a (minimal) basis for
the right, respectively left nullspace of $E$, which implies
\begin{equation}\label{wm:nullspace}
U^T E=0,\ \ E V=0.
\end{equation}
Throughout the paper we make the following assumption.
\begin{assumption} \label{wm:assumption}
The matrix $U^T A_0 V$ is nonsingular.
\end{assumption}

\medskip

The equations (\ref{wm:system}) can be separated into coupled
delay differential and delay  difference equations. When we
define
\[
\mathbf{U}= \left[{U^{\perp}}\  U\right],\ \ \mathbf{V}=\left[V^{\perp}\ V\right],\
\]
a pre-multiplication of (\ref{wm:system}) with $\UU^T$ and the substitution
\[
x=\VV\ [x_1^T\ x_2^T]^T,
\]
with $x_1(t)\in\RR^{n-\nu}$ and $x_2(t)\in\RR^{\nu}$, yield the coupled equations
{%\small
\begin{equation}\label{wm:coupled}
\begin{array}{l}
%
%\begin{eqnarray}
E^\ee \dot x_1(t)= \sum_{i=0}^m A_i^\ee x_1(t-\tau_i) +\sum_{i=0}^m A_i^\et x_2(t-\tau_i), \\
0=A_0^\tt x_2(t)+ \sum_{i=1}^m A_i^\tt x_2(t-\tau_i)
%\\
%&&
+\sum_{i=0}^m A_i^\te x_1(t-\tau_i),
\end{array}
\end{equation}
}
%
%
%\end{eqnarray}
where
\begin{equation}\label{wm:transmat2}
E^\ee= {U^{\perp}}^T E V^{\perp}
\end{equation}
and
\begin{equation}\label{wm:transmat}
\left.
\begin{array}{lll}
A_i^\ee= {U^{\perp}}^T A_i V^{\perp}, & A_i^\et= {U^{\perp}}^T A_i V,&\\
A_i^\te= {U}^T A_i V^{\perp}, & A_i^\tt= {U}^T A_i V, & i=0,\ldots,m.
\end{array}
\right.
\end{equation}
In (\ref{wm:coupled}) matrix $E^\ee$ is invertible, following from
\[
\mathrm{rank}(E^{(11)})=\mathrm{rank}(\mathbf{U}^T E\mathbf{V})=\mathrm{rank}(E)=n-\nu,
\]
and matrix $A_0^{(22)}$ is invertible as well, induced by Assumption~\ref{wm:assumption}.

%

%Equations (\ref{wm:coupled}) are semi-explicit delay differential algebraic equations of index~1,
% because  delay differential equations are obtained by differentiating the second equation. It precludes the occurrence of
% impulsive solutions~\cite{fridman}. Moreover, the invertibility of $A_0^{(22)}$
% prevents
% that the equations are of advanced type and, hence, non-causal.

\vspace{-.5cm}

\section{Spectral properties and stability}\label{wm:sectionspect}

 %In this section the spectral properties of equation (\ref{wm:system}) are discussed. In the technical derivation  connections  with the neutral equation
%obtained by differentiating the second equation in (\ref{wm:coupled}),
%play an important role.

\subsection{Exponential stability}

Stability conditions for the zero solution of (\ref{wm:system}) can be expressed in terms of the position of the \emph{characteristic roots}, i.e.,~the roots of the equation
\begin{equation}\label{wm:defdelta}
\det \Delta(\lambda)=0,
\end{equation}
where $\Delta$ is the characteristic matrix,
$
\Delta(\lambda):=\lambda E-A_0-\sum_{i=1}^m A_i e^{-\lambda\tau_i}.
$
%
%
%
% Note that the effect of the differentiation on the characteristic
%roots is restricted to (possibly) introducing characteristic roots at zero.
%
In particular, we have the following result.
\begin{proposition}\label{wm:propexp}
The null solution of (\ref{wm:system}) is exponentially stable if and only if $c<0$,
where $c$ is the \emph{spectral abscissa},
$
c:=\sup\left\{\Re(\lambda):\ \det\Delta(\lambda)=0\right\}.
$
\end{proposition}

\subsection{Continuity of the spectral abscissa and strong stability }\label{wm:parcont}
%
%We now address continuity properties of the spectral abscissa (\ref{wm:system}) with respect to the delay parameters.
We discuss the dependence of the spectral abscissa of (\ref{wm:system}) on the delay parameters $\vec\tau=(\tau_1,\ldots,\tau_m)$.
In general the function
\begin{equation}\label{wm:defc}
\vec\tau\in(\RR_{0}^+)^{m}\mapsto c(\vec\tau)
\end{equation}
is not everywhere continuous, which carries over from the spectral properties of delay
difference equations (see, e.g.,~\cite{expopo,TW-report-286,wimneutral2}). In the light of this observation we first outline properties of the function
\begin{equation}\label{wm:defcd}
\vec\tau\in(\RR_{0}^+)^{m}\mapsto c_D(\vec\tau):=\sup\left\{\Re(\lambda):\ \det \Delta_D(\lambda;\ \vec\tau)=0\right\},
\end{equation}
with
\begin{equation}\label{wm:defdeltad}
\Delta_D(\lambda;\ \vec\tau):=U^T A_0V +\sum_{i=1}^m U^T A_i V e^{-\lambda\tau_i}.
\end{equation}
Note that  (\ref{wm:defdeltad}) can be interpreted as the characteristic matrix of the delay difference equation
\begin{equation}\label{wm:asdifference}
U^T A_0V z(t) +\sum_{i=1}^m U^T A_i V z(t-\tau_i)=0,
\end{equation}
associated with the neutral equation obtained by differentiating the second equation in (\ref{wm:coupled}).

The property that the
function (\ref{wm:defcd}) is not continuous led in \cite{41471}  to the smallest upper bound, which
is `\emph{insensitive}' to small delay changes.
\begin{definition} \label{wm:defnew}
For $\vec \tau\in(\RR_{0}^+)^{m}$, let ${C_D}(\vec \tau)\in\RR$ be defined as
\[
{C_D}(\vec \tau):=\lim_{\epsilon\rightarrow 0+}
c_D^{\epsilon}(\vec\tau),
\]
where
\[
c_D^{\epsilon}(\vec \tau):=\sup
\left\{c_D(\vec \tau+\delta\vec \tau):\
\delta\vec \tau\in\RR^m\mathrm{\ and\
}\|\delta\vec \tau\|\leq\epsilon\right\}.
\]
\end {definition}
%Clearly we have ${C_D}(\vec \tau)\geq c_D(\vec \tau)$, and
%the inequality can be \emph{strict}, as shown in
%\cite{WimTom05}.
Several properties of this upper bound on $c_D$, which we call the robust spectral abscissa of the delay difference equation (\ref{wm:asdifference}), are listed below
(see \cite[Section 3]{wimdae} for an overview).
\begin{proposition}\label{wm:maintheorem}
The following assertions hold:
\begin{enumerate}
\item
the function
\[
\vec \tau\in(\RR^{+}_0)^m\mapsto{C_D}(\vec \tau)
\]
is continuous;
\item for every $\vec\tau\in(\RR_0^+)^m$, the quantity
 ${C_D}(\vec\tau)$ is
equal to the unique zero of the strictly decreasing
function
\begin{equation}\label{wm:decrease}
\zeta\in\RR\rightarrow f(\zeta;\ \vec\tau)-1,
\end{equation}
where
$f:\ \RR\rightarrow\RR^+$ is defined by
{%\small
\begin{multline}\label{wm:deff}
f(\zeta;\ \vec\tau)
:=\max_{\vec\theta\in [0,\
2\pi]^m} \rho\left(\sum_{k=1}^m (U^TA_0V)^{-1} (U^T A_k V) e^{-\zeta\tau_k}
e^{j\theta_k}\right);
\end{multline}}
\item ${C_D}(\vec \tau)=c_D(\vec \tau)$ for rationally
independent
\footnote{The $m$ components of
$\vec\tau=(\tau_1,\ldots,\tau_m)$ are rationally
independent if and only if $\sum_{k=1}^m n_k \tau_k=0,\
n_k\in\ZZ$ implies $n_k=0,\ \forall k=1,\ldots,m$.
}
%For
%instance, two delays $\tau_1$ and $\tau_2$ are rationally
%independent if their ratio is an irrational number.}
%$\vec \tau$;
%
%
\item for all $\vec \tau_1,\vec \tau_2\in(\RR_0^+)^m$, we have
\begin{equation}\label{wm:defxi}
\mathrm{sign}\left( C_D(\vec \tau_1)\right)= \mathrm{sign}\left( C_D(\vec \tau_2)\right):=\Xi;
\end{equation}
\item
 $\Xi<0$ $(>0)$ holds if and only if $\gamma_0<1\ (>1)$ holds, where
\begin{equation}\label{wm:defgamma0}
\gamma_0:=\max_{\vec\theta\in [0,\
2\pi]^m} \rho\left(\sum_{k=1}^m (U^TA_0V)^{-1} (U^T A_k V)
e^{j\theta_k}\right).
\end{equation}
\end{enumerate}
\end{proposition}
%For the single delay case, some of the expressions can be simplified.
%\begin{corollary}
%If $m=1$ then we have
%\[
%C_D(\vec\tau)=\frac{1}{\tau_1} \log\left\{ \rho((U^TA_0V)^{-1} (U^T A_1 %V))\right\}
%\]
%and
%\[
%\gamma_0=\rho((U^TA_0V)^{-1} (U^T A_1 V)).
%\]
%\end{corollary}

We now come back to the DDAE (\ref{wm:system}), more precisely, to the properties of the spectral abscissa function (\ref{wm:defc}). The following  two technical lemmas make connections
between the characteristic roots of (\ref{wm:system}) and the zeros of (\ref{wm:defdeltad}).
\begin{lemma}\label{wm:lem1}
There exists a sequence
$\{\lambda_k\}_{k\geq 1}$ of characteristic roots of (\ref{wm:system}) satisfying
\[
\lim_{k\rightarrow\infty}\Re(\lambda_k)=c_D,\ \  \lim_{k\rightarrow\infty}\Im(\lambda_k)=\infty.
\]
%Consequently,
%\[
%c(\vec\tau)\geq c_D(\vec\tau),\ \forall\vec\tau\in\left(\RR_0^+\right)^m.
%\]
\end{lemma}
%
%
%\begin{lemma}\label{wm:lem2}
%For every $\epsilon>0$ and $\vec\tau\in\left(\RR_0^+\right)^m$, the equation %(\ref{wm:system})
%only has a finite number of
%characteristic roots  in the half plane described by
%\[
%\left\{\lambda\in\CC:\ \Re(\lambda)\geq{C_D}(\vec
%\tau)+\epsilon\right\}.
%\]
%
%\end{lemma}

\begin{lemma}\label{wm:lem2}
For every $\epsilon>0$ the number of characteristic roots of (\ref{wm:system}) in the half plane
\begin{equation}\label{wm:halfplaneeps}
\left\{\lambda\in\CC:\ \Re(\lambda)\geq{C_D}(\vec
\tau)+\epsilon\right\}
\end{equation}
is finite.
\end{lemma}

The lack of continuity of the spectral abscissa function (\ref{wm:defc}) leads us again to an upper bound that takes into account the effect of small delay perturbations.
\begin{definition}
For $\vec \tau\in(\RR_{0}^+)^{m}$, let the \emph{robust spectral abscissa} ${C}(\vec \tau)$ of (\ref{wm:system}) be defined as
\begin{equation}\label{wm:defC}
{C}(\vec \tau):=\lim_{\epsilon\rightarrow 0+}
c^{\epsilon}(\vec\tau),
\end{equation}
where
\[
c^{\epsilon}(\vec \tau):=\sup
\left\{c(\vec \tau+\delta\vec \tau):\
\delta\vec \tau\in\RR^m\mathrm{\ and\
}\|\delta\vec \tau\|\leq\epsilon\right\}.
\]
\end {definition}
The following characterization of the robust spectral abscissa (\ref{wm:defC}) constitutes the main result of this section. Its proof can be found in \cite{wimdae}.
\begin{proposition}
The following assertions hold:
\begin{enumerate}
\item
the function
\begin{equation}\label{wm:defc2}
\vec \tau\in(\RR^{+}_0)^m\mapsto{C}(\vec \tau)
\end{equation}
is continuous;
\item for every $\vec\tau\in(\RR_0^+)^m$, we have
\begin{equation}\label{wm:defccd}
C(\vec\tau)=\max({C_D}({\vec\tau}), c({\vec\tau})).
\end{equation}
\end{enumerate}
\end{proposition}

In line with the sensitivity of the spectral abscissa with respect to infinitesimal delay perturbations, which has been resolved by considering the robust spectral abscissa (\ref{wm:defC}) instead, we define the concept of strong stability\footnote{This
terminology is borrowed from the theory of neutral delay differential equations~\cite{have:02,41471}.}.
\begin {definition}
The null solution of  (\ref{wm:system}) is strongly
exponentially stable if there exists a number $\hat \tau>0$
such that the null solution of
\[
E\dot x(t)=A_0+\sum_{k=1}^{m} A_k
x(t-(\tau_k+\delta\tau_k))
\]
is exponentially stable for all $\delta\vec
\tau\in(\RR^+)^m$
satisfying $\|\delta\vec \tau\|<\hat \tau$ and $\tau_k+\delta \tau_k\geq 0,\ k=1,\ldots,m$.
\end{definition}
The following result provides necessary and sufficient conditions
for exponential stability.
\begin {theorem} \label{wm:strongneut}
The null solution of  (\ref{wm:system})
is strongly exponentially stable if and only if $C(\vec\tau)<0$, or, equivalently,  $c(\vec\tau)<0$ and $\gamma_0<1$, where $\gamma_0$ is defined by~(\ref{wm:defgamma0}).
\end{theorem}

\vspace{-.75cm}

\section{Robust stabilization by eigenvalue optimization}\label{wm:sectionstabil}
We now consider the equations
\begin{equation}\label{wm:systemp}
E \dot x(t)=A_0(\vec p) x(t)+\sum_{i=1}^m A_i(\vec p) x(t-\tau_i),
\end{equation}
where the system matrices linearly depend on parameters $\vec p\in\mathbb{R}^{n_p}$.  In control applications these parameter usually correspond to controller parameters. For example, in the feedback interconnection (\ref{wm:motex1})-(\ref{wm:motex2}) they may arise from a parameterization of the matrices $(\hat F_i,\hat G_i, \hat H_i, \hat L_i)$.
%
%
%In \S\ref{wm:partwoopt} the stabilization problem for (\ref{wm:systemp}) is related to two optimization problems and in \S\ref{wm:algstab} the corresponding optimization algorithms are briefly discussed.

To impose exponential stability of the null solution of (\ref{wm:systemp}) it is necessary to find values of $\vec p$ for which the spectral abscissa is strictly negative. If the achieved stability is required to be robust against  small delay perturbations, this requirement must be strengthened to the negativeness of the robust spectral abscissa.  This brings us to the optimization problem
\begin{equation}\label{wm:optproblem1}
\inf_{\vec p} C(\vec\tau;\ \vec p).
\end{equation}
Strongly stabilizing values of $\vec p$ exist if the objective function can be made strictly negative.  By Theorem~\ref{wm:strongneut} the latter can be evaluated as
\begin{equation}\label{wm:dichot}
C(\vec\tau;\ \vec p)=\max(c(\vec\tau;\ \vec p), C_D(\vec\tau;\ \vec p)).
\end{equation}

An alternative approach  consists of solving the constrained optimization problem
\begin{equation}\label{wm:optproblem2}
\begin{array}{c}
\inf_{\vec p} c(\vec\tau;\ \vec p),
 \mathrm{\ subject\ to}\ \
   \gamma_0(\vec p)<\gamma,
\end{array}
\end{equation}
where $\gamma<1$.  If the objective function is strictly negative, then the satisfaction of the constraint implies strong stability. Problem (\ref{wm:optproblem2}) can be solved using the barrier method proposed in \cite{tomasIMA}, which is on its turn inspired by interior point methods, see, e.g., \cite{boydvandenberghe}.
  The first step consists of finding a feasible point, i.e.,  a set of values for $\vec p$ satisfying the constraint. If the
feasible set is nonempty such a point can be found by solving
\begin{equation}\label{wm:optprob2}
\min_{\vec p} \gamma_0(\vec p).
\end{equation}
Once a feasible point $\vec p=\vec p_0$ has been obtained one can solve in the next step the unconstrained optimization
problem
\begin{equation}\label{wm:repeated}
\min_{\vec p} \left\{c(\vec p)-r\log(\gamma-\gamma_0(\vec p))\right\}
\end{equation}
where $r > 0$ is a small number and $\gamma$ satisfies
\[
\gamma_0(\vec p)<\gamma\leq 1.
\]
%$\gamma\in (\gamma_0(\vec p_0),\ 1]$.
The second term (the barrier) assures that the feasible set
cannot be left when the objective function is decreased in a quasi-continuous way (because the objective
function will go to infinity when $\gamma_0\rightarrow\gamma$).
If (\ref{wm:repeated}) is repeatedly solved for decreasing values of $r$ and
with the previous solution as starting value, a solution of (\ref{wm:optproblem2}) is obtained.
 %Strong \emph{exponential} stability can be imposed by setting $\gamma$ strictly smaller than one.
%
%
%

 In \cite{wimdae} it has been shown that the objective functions for the optimization problem (\ref{wm:optproblem1}) and for the subproblems (\ref{wm:optprob2}) and (\ref{wm:repeated}) are in general not everywhere differentiable. They might even be not everywhere Lipschitz continuous, yet they are differentiable almost everywhere. These properties preclude the use of standard optimization methods, developed for smooth problems.  Instead we use
a combination of BFGS with weak Wolfe line search and gradient sampling, as implemented in the MATLAB code HANSO~\cite{overtonhanso}. The overall algorithm only requires the evaluation of the objective function, as well as its derivatives with respect to the controller parameters, \emph{whenever} it is differentiable. The spectral abscissa can be computed using a spectral discretization followed by Newton corrections. The quantities $C_D$ and $\gamma_0$ can be computed using the characterizations in Theorem~\ref{wm:maintheorem}, where the (global) maximization problems in (\ref{wm:deff}) and (\ref{wm:defgamma0}) are discretized, followed by local corrections. In all cases derivatives can be obtained from the sensitivity of individual eigenvalues with respect to the free parameters. For more details and expressions we refer to \cite{wimdae}.

\section{Illustration of the software}\label{wm:sectionex}

A MATLAB implementation of the robust stabilization algorithms is available
from
{\small
\begin{verbatim}
http://twr.cs.kuleuven.be/research/software/delay-control/stabilization/.
\end{verbatim}
}
\noindent Installation instructions can be found in the corresponding README file.

As a first example we take the system with input delay from~\cite{joris}:
\begin{equation}\label{wm:eqnumvb1}
\dot{x}(t)=A x(t) +Bu(t-\tau),\ \ \ y(t)=x(t),
\end{equation}
where
\begin{equation}\label{wm:eqnumvb2}
\begin{array}{cc}
A=\left[\begin{array}{rrr}
-0.08 &-0.03  &0.2\\
0.2   &-0.04  &-0.005\\
-0.06 &0.2    &-0.07
\end{array}\right],\ \ \ \
B=\left[\begin{array}{r}
-0.1\\
-0.2\\
0.1
\end{array}\right],\ \ \ \
\tau=5.
\end{array}
\end{equation}
We start by defining the system:
{\small
\begin{verbatim}
>> A = [-0.08 -0.03 0.2;0.2 -0.04 -0.005;-0.06 0.2 -0.07];
>> B = [-0.1;-0.2;0.1];
>> C = eye(3);
>> plant1 = tds_create({A},0,{B},5,{C},0);
\end{verbatim}
}
\noindent The uncontrolled system is unstable.
{\small \begin{verbatim}
>> max(real(eig(A)))

ans =

    0.1081
\end{verbatim}}

\begin{figure}
 \begin{center}
   \resizebox{7.0cm}{6cm}{\includegraphics{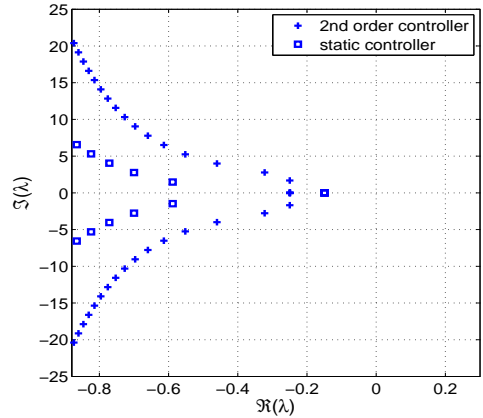}}
        \caption{Characteristic roots of the first example (\ref{wm:eqnumvb1}) and (\ref{wm:dynfeedback}), corresponding to a minimum of the spectral abscissa, for a static controller (boxes)  and a second order controller (pluses).    \label{wm:acfig}}
    \end{center}
\end{figure}

\noindent We design a stabilizing dynamic controller of the form
\begin{equation}\label{wm:dynfeedback}
\left\{\begin{array}{lll}
\dot x_c(t)&=& A_c x_c(t)+B_c y(t),\\
 u(t) &=& C_c x_c(t)+D_c y(t),\ \ \ x_c(t)\in\RR^{n_c},
\end{array}\right.
\end{equation}
using the approach of Section~\ref{wm:sectionstabil},
where we set
$
p=\mathrm{vec}\left[\begin{array}{cc} A_c & B_c \\ C_c & D_c \end{array}\right].
$
Since the transfer function from $u$ to $y$ is strictly proper, the robust spectral abscissa equals the spectral abscissa, and the optimization problems (\ref{wm:optproblem1}) and (\ref{wm:optproblem2}) reduce to the (unconstrained) minimization of the spectral abscissa.  In order to compute a controller we first specify its order,
{\small \begin{verbatim}
>> controller_order = 2;
\end{verbatim}}
\noindent and call a routine to minimize the robust spectral abscissa
{\small \begin{verbatim}
>> [controller1,f1] = stabilization_max(plant1,controller_order);
\end{verbatim}}
\noindent The optimized robust spectral abscissa and corresponding controller are given by:
{\small \begin{verbatim}
controller1 =

       E: {[2x2 double]}
      hE: 0
       A: {[2x2 double]}
      hA: 0
      B1: {[2x3 double]}
     hB1: 0
      C1: {[0.2098 0.9492]}
     hC1: 0
     D11: {[0.8826 1.1548 0.6538]}
    hD11: 0
\end{verbatim}}
\noindent where empty fields of the controller are omitted for space considerations.
{\small
\begin{verbatim}
f1 =

 -0.2496
\end{verbatim}}
\noindent We define the closed-loop system
{\small \begin{verbatim}
>> clp1 = closedloop(plant1,controller1);
\end{verbatim}}
\noindent and compute its rightmost characteristic roots (where the "l1"-field refers to the application of Newton corrections):
{\small \begin{verbatim}
>> options = tdsrootsoptions;
>> eigenvalues1 = compute_roots_DDAE(clp1,options);
>> eigenvalues1.l1.'

ans =

  -0.2496 + 0.0251i   -0.2496 - 0.0251i
\end{verbatim}}
\noindent
We can compute all eigenvalues with real part larger than $-0.9$ by the following code, which leads to $37$ returned eigenvalues.
{\small
\begin{verbatim}
>> options.minimal_real_part = -0.9;
>> eigenvalues1 = compute_roots_DDAE(clp1,options);
>> size(eigenvalues1.l1)

ans =

    37     1
\end{verbatim}
}
\noindent
We plot the closed-loop characteristic roots.
{\small
\begin{verbatim}
>> p1 = eigenvalues1.l1; plot(real(p1),imag(p1),'+');
\end{verbatim}
}
\noindent We now repeat the computations for a static controller:
{\small
\begin{verbatim}
>> controller_order = 0;
>> [controller2,f2] = stabilization_max(plant1,controller_order);
\end{verbatim}
}
\noindent and add the optimized spectrum to our plot:
{\small
\begin{verbatim}
>> clp2 = closedloop(plant1,controller2);
>> eigenvalues2 = compute_roots_DDAE(clp2,options);
>> p2 = eigenvalues2.l1; hold on; plot(real(p2),imag(p2),'s');
\end{verbatim}
}
\noindent The result is displayed in Figure~\ref{wm:acfig}. Note that the extra degrees of freedom in the dynamic controller lead to a further reduction of the spectral abscissa.

\begin{figure}
 \begin{center}
 \resizebox{5.8cm}{!}{\includegraphics{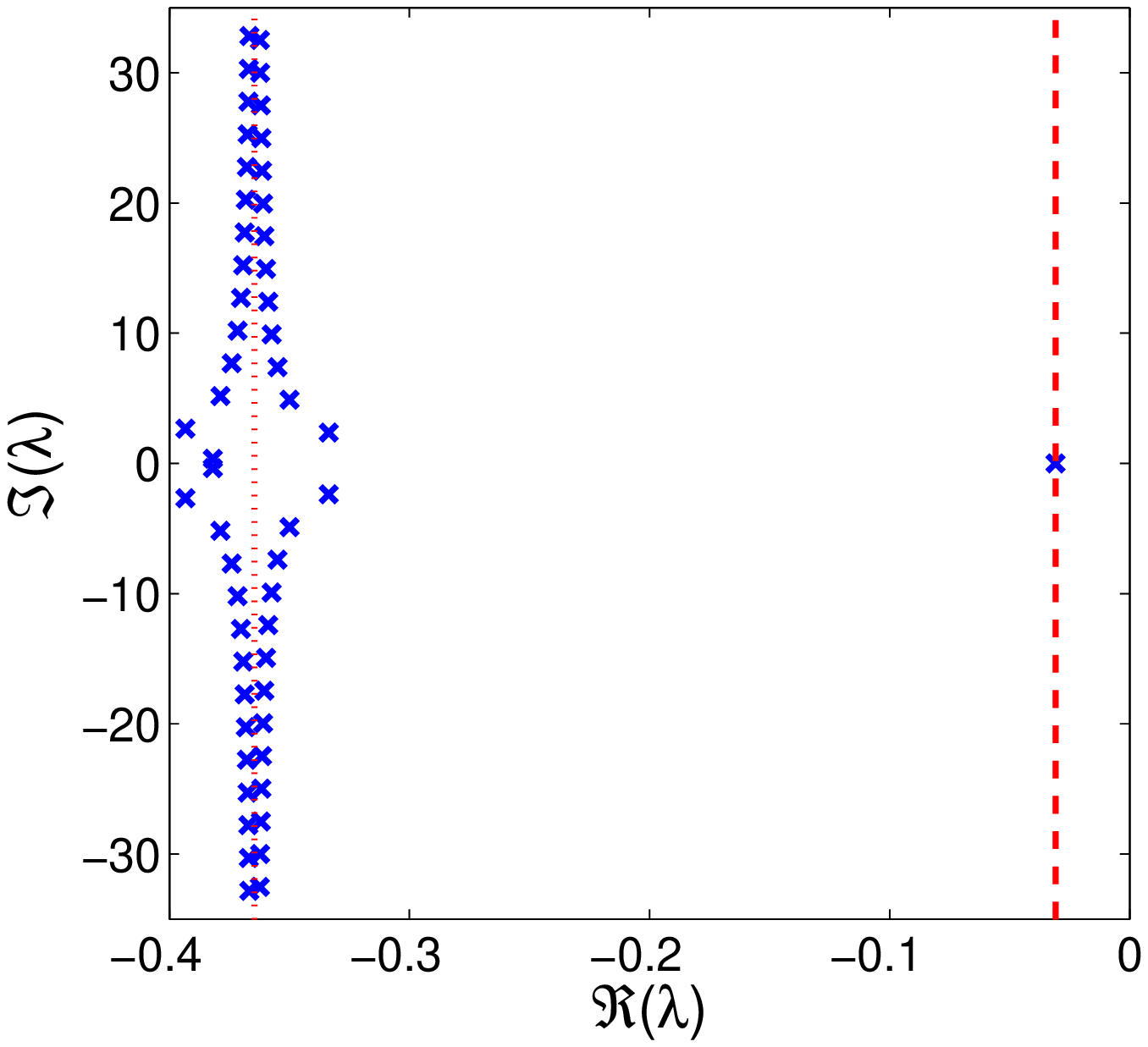}}
   \resizebox{5.8cm}{!}{\includegraphics{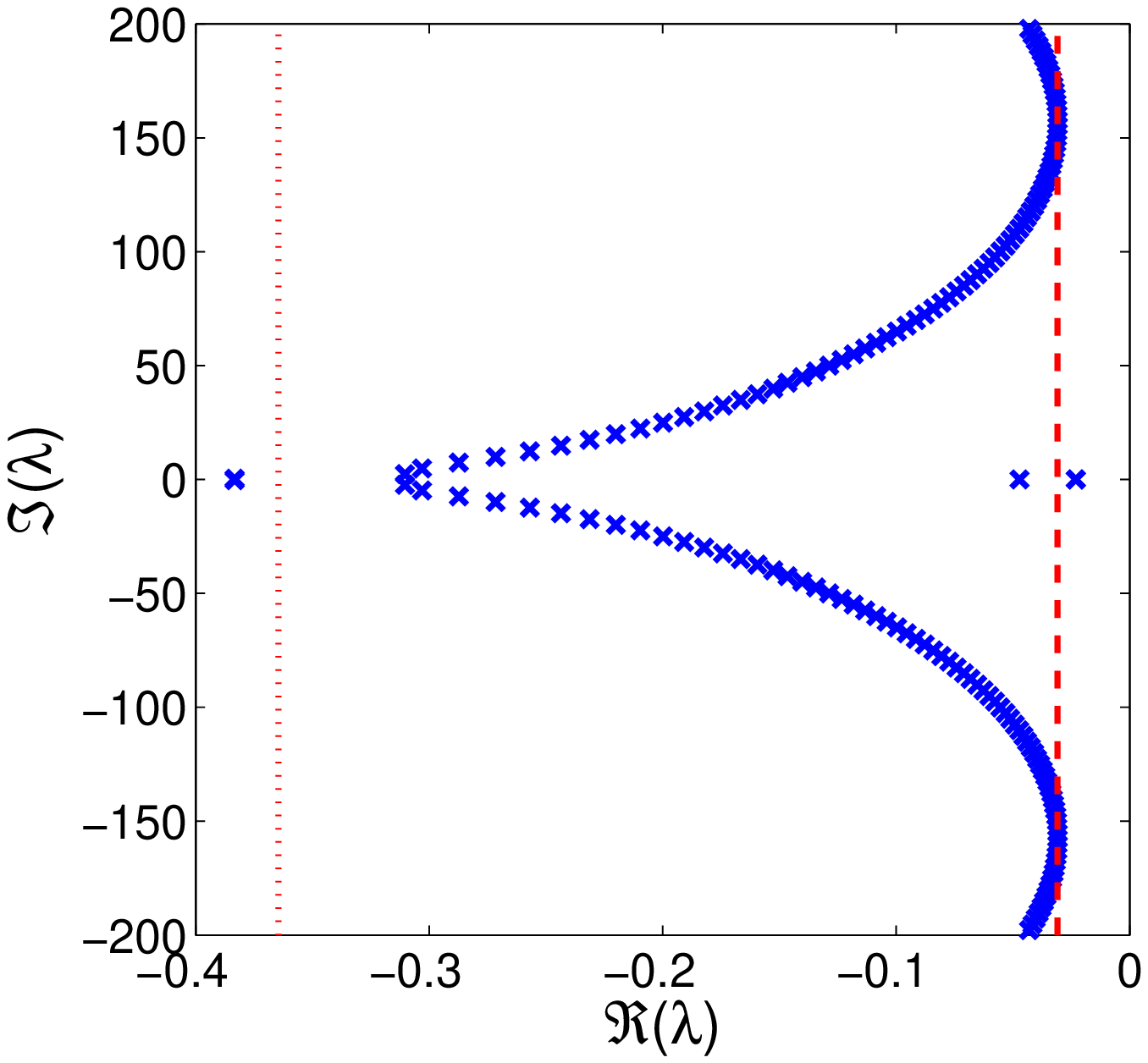}}
        \caption{ (left) Characteristic roots corresponding to the minimum of the robust spectral abscissa of  the second example (\ref{wm:eqnumvb1}) and (\ref{wm:newoutput}), using a static controller. The rightmost characteristic roots, $\lambda\approx-0.0309$, has multiplicity three. (right) Effect on the characteristic roots of a perturbation of the delays $(2.5,5)$ in (\ref{wm:newoutput}) to~$(2.51,5)$.         \label{wm:figtedoen}}
    \end{center}
\end{figure}

For the second example we assume that the measured output of  system (\ref{wm:eqnumvb1}) is instead given by
\begin{equation}\label{wm:newoutput}
\tilde y(t)= x(t) +
\left[\begin{array}{ccc}
3 &  4 & 1
\end{array}\right]^T u(t-2.5)
+
\left[\begin{array}{rrr}
2/5 & -2/5 & - 2/5
\end{array}\right]^T u(t-5).
\end{equation}
\noindent The difference with the previous example is that there are two feedthrough terms which are both delayed. We define the plant object
{\small \begin{verbatim}
>> plant2 = setfield(plant1,'D11',{[3;4;1],[2/5;-2/5;-2/5]});
>> plant2.hD11 = [2.5 5];
\end{verbatim}}
\noindent Once again we design a static controller, $u(t)=D_c\tilde y(t)$.
%\begin{equation}\label{wm:staticvb2}
%u(t)= D_c \tilde y(t).
%\end{equation}
In this case there is a high-frequency path in the control loop.
Solving the optimization problem (\ref{wm:optproblem1}) leads to
\begin{equation}\label{wm:Copt}
C=-0.0309,\ \ \ D_c=[0.0409\ \ 0.0612\ \ 0.3837],
\end{equation}
as can be seen from
{\small \begin{verbatim}
>> [controller1,f1] = stabilization_max(plant2,controller_order);
>> f1

f1 =

   -0.0309
>> controller1.D11{1}

ans =

    0.0409    0.0612    0.3837
   \end{verbatim}}
We compute the rightmost characteristic roots of the closed-loop system.
{\small\begin{verbatim}
>> clp1 = closedloop(plant2,controller1);
>> eigenvalues1 = compute_roots_DDAE(clp1,options);
Warning: case C_D>=c.
Spectral discretization with 16 points (lowered if maximum size of
the eigenvalue problem is exceeded)
N= 15
>> eigenvalues1.l1.'

ans =

  -0.3740 + 7.6893i  -0.3740 - 7.6893i  -0.3788 + 5.1779i  -0.3788 - 5.1779i
  -0.3499 + 4.8863i  -0.3499 - 4.8863i  -0.3934 + 2.6712i  -0.3934 - 2.6712i
  -0.3336 + 2.3789i  -0.3336 - 2.3789i  -0.0309            -0.0309 + 0.0001i
  -0.0309 - 0.0001i  -0.3819 + 0.3603i  -0.3819 - 0.3603i
\end{verbatim}}
\noindent We conclude that the optimum is characterized by three rightmost characteristic roots. This might sound counter-intuitive because the number of degrees of freedom in the controller is also three. The explanation is related to the issued warning: we are in a situation where $C_D\geq c$. In fact the optimum of (\ref{wm:optproblem1}) is characterized by an equality between $C_D$ and the spectral abscissa $c$, the latter corresponding to a rightmost root with multiplicity three.   To illustrate this, we have recomputed the characteristic roots where we set $N$, the number of discretization points in the spectral method, to a high number in such a way that the high-frequency roots are captured. In the left pane of Figure~\ref{wm:figtedoen} we show the rightmost characteristic roots corresponding to the minimum of the robust spectral abscissa (\ref{wm:Copt}). The dotted line corresponds to $\Re(\lambda)=c_D$, the dashed line to $\Re(\lambda)=C_D$.  In order to illustrate that we indeed have $c=C_D$ we depict in the right pane of Figure~\ref{wm:figtedoen} the rightmost characteristic roots after perturbing the delay value $2.5$ in
(\ref{wm:newoutput}) to $2.51$.

\medskip

With our software we can also solve the constrained optimization problem~(\ref{wm:optproblem2}). With the default parameters $r=10^{-3}$ and $\gamma=1-10^{-3}$ in the relaxation~(\ref{wm:repeated}) we get the following result:
{\small \begin{verbatim}
>> [controller2,f2] = stabilization_barrier(plant2,controller_order);
>> controller2.D11{1}

ans =

    0.0249    0.1076    0.3173
>> clp2 = closedloop(plant2,controller2);
>> eigenvalues2=compute_roots_DDAE(clp2,options);
Warning: case C_D>=c.
Spectral discretization with 16 points (lowered if maximum size of
the eigenvalue problem is exceeded)
N= 15
>> max(real(eigenvalues2.l1))
ans =

  -0.0345
\end{verbatim}}
Compared to (\ref{wm:Copt}), where we had $C=c=C_D$, a further reduction of the spectral abscissa
to $c=-0.0345$ has been achieved, at the price of an increased value of $C_D$ (equal to $-0.00602$). This is expected because the constraint $\gamma_0<1$ imposes robustness of stability, yet no bound on the exponential decay rate of the solutions.

\vspace{-.5cm}

\vspace{-1.5cm}

\section{Duality with the $\mathcal{H}_{\infty}$ problem}

In a practical control design the stabilization phase is usually only a first step in the overall design procedure.  Consider now the (subsequent) fixed-order $\mathcal{H}_{\infty}$ synthesis problem, where the aim is to optimize the $\mathcal{H}_{\infty}$  norm of
\[
\begin{array}{l}
G(\lambda)=C(\lambda E-A_0-\sum_{i=1}^m A_i e^{-\lambda\tau_i})^{-1} B
\end{array}
\]
as a function of parameters on which the system matrices depend.

It turns out the function $\vec\tau\mapsto \|G(j\omega;\ \vec\tau)\|_{\mathcal{H}_{\infty}}$ has a very similar behavior to the spectral abscissa function (\ref{wm:defc}). In particular it is not everywhere continuous. Moreover, the discontinuities are all related to the behavior of the transfer function at large frequencies (analogous to the behavior of eigenvalues with large imaginary parts in \S\ref{wm:parcont}). This high frequency behavior is described by the associated \emph{asymptotic transfer function}
\[
G_a(\lambda) :=-C V (U^T A_0 V +\sum_{i=1}^m U^T A_i V e^{-\lambda\tau_i})^{-1} U^TB,
\]
which takes the role of the associated delay-difference equation (\ref{wm:asdifference}). Finally, the sensitivity w.r.t.~small delay perturbations leads to the definition of the \emph{strong} $\mathcal{H}_{\infty}$ norm (analogous to strong stability), defined as:
\begin{multline} \nonumber \hspace{-.3cm}
   \interleave G(j\omega;\ \vec \tau)\interleave_{\mathcal{H}_\infty}:=
   \lim_{\epsilon\rightarrow 0+}
\sup \{\|G(j\omega;\ \vec \tau+\delta\vec\tau)\|_{\mathcal{H}_{\infty}}: \delta\vec\tau\in\left(\mathbb{R}^+\right)^m,\ \|\delta\vec\tau\|_2<\epsilon\}.
\end{multline}
 The computation of the strong $\mathcal{H}_{\infty}$ norm  involves a tradeoff between the behavior of the transfer function at small and large frequencies, similar to the result of Theorem~\ref{wm:strongneut} on strong stability, and it can be optimized using the same algorithms.
For the details, we refer to the article~\cite{reporthinf} and to the corresponding software available at
{\small
\begin{verbatim}
http://twr.cs.kuleuven.be/research/software/delay-control/hinfopt/.
\end{verbatim}}

\vspace{-1cm}

{\small
\section*{Acknowledgements} \vspace{-.5cm}
This work has been supported by the Programme of Interuniversity Attraction
Poles of the Belgian Federal Science Policy Office (IAP P6- DYSCO), by
OPTEC, the Optimization in Engineering Center of the K.U.Leuven, by the project STRT1-
09/33 of the K.U.Leuven Research Council and the project G.0712.11N of
the Research Foundation - Flanders (FWO).}

\vspace{-.75cm}

% FINAL: inline reference list


\begin{thebibliography}{99.}

\bibitem{expopo}
C.E. Avellar and J.K. Hale.
\newblock On the zeros of exponential polynomials.
\newblock {\em Mathematical analysis and applications}, 73:434--452, 1980.

\bibitem{boydvandenberghe}
S.~Boyd and L.~Vandenberghe.
\newblock {\em Convex optimization}.
\newblock Cambridge University Press, 2004.

\bibitem{fridman}
E.~Fridman and U.~Shaked.
\newblock {$H_{\infty}$-control of linear state-delay descriptor systems: an
  LMI approach}.
\newblock {\em Linear Algebra and its Applications}, 351-352:271--302, 2002.

\bibitem{reporthinf}
S.~Gumussoy and W.~Michiels.
\newblock Fixed-order H-infinity control for interconnected systems using delay
  differential algebraic equations.
\newblock {\em {SIAM} Journal on Control and Optimization}, 49(5):2212--2238,
  2011.

\bibitem{have:02}
J.K. Hale and S.M Verduyn~Lunel.
\newblock Strong stabilization of neutral functional differential equations.
\newblock {\em IMA Journal of Mathematical Control and Information}, 19:5--23,
  2002.

\bibitem{41471}
W.~Michiels and T.~Vyhlidal.
\newblock {A}n eigenvalue based approach to the robust stabilization of linear
  time-delay systems of neutral type.
\newblock {\em Automatica}, 41(6):991--998, 2005.

\bibitem{MichielsIFACTDS2012}
Michiels, W. and Gumussoy, S.
\newblock Eigenvalue based Analysis and Controller Synthesis for Systems described by Delay Differential Algebraic Equations.
\newblock {\em 10th IFAC Workshop on Time Delay Systems}, June 22-24, Northeastern University, USA, 2012. \newblock {\em IFAC-PapersOnLine}, 144--149, doi: 10.3182/20120622-3-US-4021.00015, 2012.

\bibitem{TW-report-286}
W.~Michiels, K.~Engelborghs, D.~Roose, and D.~Dochain.
\newblock Sensitivity to infinitesimal delays in neutral equations.
\newblock {\em {SIAM} Journal on Control and Optimization}, 40(4):1134--1158,
  2002.

\bibitem{wimdae}
W.~Michiels.
\newblock Spectrum based stability analysis and stabilization of systems
  described by delay differential algebraic equations.
\newblock {\em {IET} Control Theory and Applications}, 5(16):1829--1842, 2011.

\bibitem{wimneutral2}
W.~Michiels, T.~Vyhl\'idal, P.~Zit\'ek, H.~Nijmeijer, and D.~Henrion.
\newblock Strong stability of neutral equations with an arbitrary delay
  dependency structure.
\newblock {\em {SIAM} Journal on Control and Optimization}, 48(2):763--786,
  2009.

\bibitem{overtonhanso}
M.~Overton.
\newblock {HANSO}: a hybrid algorithm for nonsmooth optimization.
\newblock Available from \verb|http://cs.nyu.edu/overton/software/hanso/|,
  2009.

\bibitem{joris}
J.~Vanbiervliet, B.~Vandereycken, W.~Michiels, and S.~Vandewalle.
\newblock A nonsmooth optimization approach for the stabilization of time-delay
  systems.
\newblock {\em ESAIM Control, Optimisation and Calculus of Variations},
  14(3):478--493, 2008.

\bibitem{tomasIMA}
T.~Vyhlidal, W.~Michiels, and P.~McGahan.
\newblock Synthesis of strongly stable state-derivative controllers for a time
  delay system using constrained non-smooth optimization.
\newblock {\em {IMA} Journal of Mathematical Control and Information},
  27(4):437--455, 2010.


\end{thebibliography}
\end{document}